\documentclass[aps, twocolumn, nobibnotes, superscriptaddress, notitlepage, nobalancelastpage, noeprint, prb]{revtex4-2}

\usepackage[usenames,dvipsnames]{xcolor}
\usepackage[linktoc=page,colorlinks,urlcolor=RedViolet,citecolor=RedViolet,linkcolor=RedViolet]{hyperref}
\usepackage{float}
\usepackage{graphicx}
\usepackage{mathrsfs}
\usepackage{overpic}
\usepackage{wrapfig}
\usepackage{braket}
\usepackage{color}
\usepackage{verbatim}
\usepackage{amssymb,amsmath,mathtools}
\usepackage{upgreek}
\usepackage{bbold}
\usepackage{cancel}
\usepackage{textgreek}
\usepackage[normalem]{ulem}

\definecolor{darkblue}{rgb}{0.0 0.0 0.78}
\definecolor{darkred}{rgb}{0.5 0.0 0.0}

\newcommand{\UMDphy}{Department of Physics, University of Maryland, College Park, Maryland 20742, USA}
\newcommand{\QTC}{Quantum Technology Center, University of Maryland, College Park, Maryland 20742, USA}
\newcommand{\UMDEECS}{Department of Electrical Engineering and Computer Science,
University of Maryland, College Park, Maryland 20742, USA}

\newcommand{\UMDBioE}{Fischell Department of Bioengineering, University of Maryland, College Park, Maryland 20742, USA}

\begin{document}

\title{Quantum Diamond Microscope for Narrowband Magnetic Imaging with High Spatial and Spectral Resolution}
\date{\today}

\author{Zechuan Yin}
\affiliation{\UMDEECS}
\affiliation{\QTC}

\author{Jiashen Tang}
\affiliation{\QTC}
\affiliation{\UMDphy}

\author{Connor A. Hart}
\affiliation{\UMDEECS}
\affiliation{\QTC}

\author{John W. Blanchard}
\affiliation{\UMDEECS}
\affiliation{\QTC}

\author{Xinyan Xiang}
\affiliation{\QTC}

\author{Saipriya Satyajit}
\affiliation{\QTC}
\affiliation{\UMDphy}

\author{Smriti Bhalerao}
\affiliation{\QTC}
\affiliation{\UMDBioE}

\author{Tao Tao}
\affiliation{\QTC}
\affiliation{\UMDphy}

\author{Stephen J. DeVience}
\affiliation{\QTC}

\author{Ronald L. Walsworth}
\email{walsworth@umd.edu}
\affiliation{\UMDEECS}
\affiliation{\QTC}
\affiliation{\UMDphy}

\begin{abstract}
The quantum diamond microscope (QDM) is a recently developed technology for near-field imaging of magnetic fields with micron-scale spatial resolution, under ambient conditions. 
In the present work, we integrate a QDM with a narrowband measurement protocol and a lock-in camera; and demonstrate imaging of radiofrequency (RF) magnetic field patterns produced by microcoils, with spectral resolution $\approx1$\,Hz.  
This RF-QDM provides multi-frequency imaging with a central detection frequency that is easily tunable over the MHz-scale, allowing spatial discrimination of both crowded spectral peaks and spectrally well-separated signals.  
The present instrument has spatial resolution $\approx2\,\mathrm{\mu m}$, field-of-view $\approx300\times300\,\mathrm{\mu m^2}$, and per-pixel sensitivity to narrowband fields $\sim{1}\,$nT$\cdot$Hz$^{-1/2}$.  
Spatial noise can be reduced to the picotesla scale by signal averaging and/or spatial binning.  
The RF-QDM enables simultaneous imaging of the amplitude, frequency, and phase of narrowband magnetic field patterns at the micron-scale, with potential applications in real-space NMR imaging, AC susceptibility mapping, impedance tomography, analysis of electronic circuits, and spatial eddy-current-based inspection.

\end{abstract}

\maketitle

\section{Introduction}

\begin{figure}
    \centering
    \includegraphics[width=\linewidth]{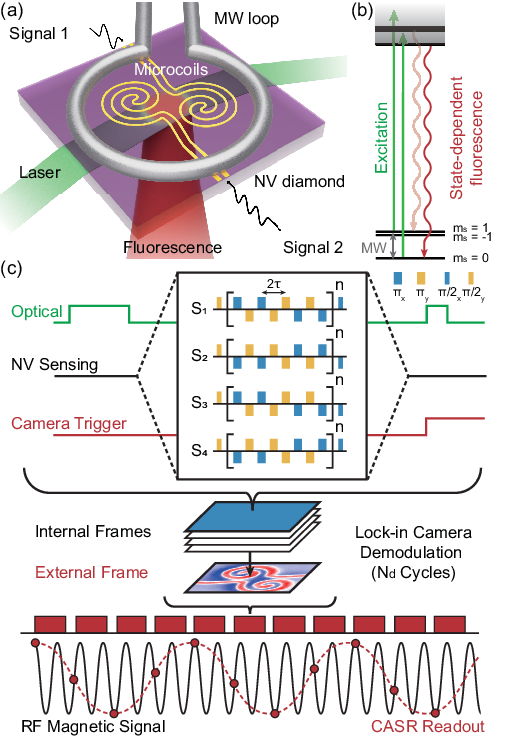}
    \caption{(a) Widefield imaging of near-field patterns of narrowband magnetic fields utilizing a quantum diamond microscope (QDM) operating under ambient conditions. 
    RF test signals are applied to microcoils to produce distinct magnetic field patterns across the NV-diamond surface.
    Laser and MW fields are employed for NV electronic spin manipulation, with spin state population readout through fluorescence detection.
    (b) NV energy-level diagram. The NV$^-$ center in diamond exhibits spin-dependent fluorescence with triplet ground-state spin transitions. 
    (c) RF magnetic field imaging protocol with a lock-in camera.
    Within one camera demodulation cycle, four exposures occur, with phase-modulated XY8-n sequences applied to NV spins to generate internal I\,\&\,Q frames.
    After $N_d$ cycles (i.e., $2N_d$ total internal frames), an external frame of the RF magnetic field image is obtained.
    The CASR protocol comprises interspersed blocks of single external frame measurements (red boxes) with a frame rate of $f_\mathrm{SR}$.
    The RF magnetic signal (black curve) with frequency $f_s$ is sampled via CASR readout (red dashed curve) at an alias frequency $f_s - mf_\mathrm{SR}$, where $m$ is the integer closest to $f_s/f_\mathrm{SR}$.
    } 
    \label{fig:1}
\end{figure}

Nitrogen-vacancy (NV) centers in diamond are robust and sensitive detectors of magnetic fields, among other physical quantities~\cite{Taylor2008,Barry2020}.
In particular, the quantum diamond microscope (QDM), utilizing a dense, micron-scale near-surface layer of NV centers in a diamond substrate, is a versatile platform for near-field magnetic field imaging, with diverse applications across the physical and life sciences \cite{Levine2019}.  
To date, QDMs have mostly been used to image patterns of static or broadband magnetic fields, though narrowband measurements have also been performed, e.g., for real-space NMR imaging \cite{Devience2015}, AC magnetic susceptibility mapping \cite{Dasika2023}, eddy current imaging \cite{Chatzidrosos2019}, and phase-sensitive imaging of a narrowband signal produced by stripline wire \cite{Mizuno2020}.
However, the CCD-camera-based measurements used in most QDMs to date have typically been slow due to low frame rate and long exposure time, compared to single-channel measurements using a photodiode and digitizer, restricting the measurement bandwidth and temporal resolution. 
Moreover, the long exposure time of a CCD camera leads to low NV spin-state optical contrast and a limited ability to mitigate broadband laser intensity noise, both of which result in reduced magnetic signal sensitivity \cite{Barry2020}.
Fortunately, the development of QDMs utilizing high-speed lock-in cameras provides a platform to acquire magnetic images at a faster rate, while applying phase modulation to increase the signal-to-noise ratio (SNR). 
In a recent demonstration \cite{Tang2023}, broadband magnetic imaging using Ramsey interferometry and a lock-in camera was realized, with sub-millisecond temporal resolution, micron-scale spatial resolution, and nanotesla-scale per-pixel sensitivity. 

Here, we demonstrate narrowband magnetic imaging protocols, integrated with a QDM using a lock-in camera, and capable of high spectral and spatial resolution measurements of patterns of near-field radiofrequency (RF) signals.
Beginning with dynamical decoupling sequences, we image in-phase, multi-frequency sub-nanotesla RF signals ($\sim{1}$\,MHz) produced by spiral-shaped microcoils [Fig. \ref{fig:1}(a)]; and find good agreement with numerical simulations. 
This RF-QDM operates under ambient conditions and provides lateral spatial resolution $\approx2\,\mathrm{\mu m}$, field-of-view (FOV) $\approx300\times300\,\mathrm{\mu m^2}$, and per-pixel sensitivity to narrowband fields $\sim{1}\,$nT$\cdot$Hz$^{-1/2}$.  
The central detection frequency can be easily tuned over several orders of magnitude by simple adjustments to the NV measurement protocol. 
We also show that RF-QDM measurements have good temporal stability by averaging down the spatial magnetic noise background to a few picotesla after 1 hour.
Subsequently, we apply the coherently averaged synchronous readout (CASR) \cite{Glenn2018} protocol, enabling simultaneous acquisition of amplitude and phase images of multi-frequency RF signals with a 250\,Hz spectral window and single-Hz resolution.
Given the current performance of the RF-QDM, we evaluate its feasibility for real-space NV-NMR imaging of samples placed on the diamond imager chip, estimating the required acquisition times for imaging both thermally polarized and hyperpolarized NMR signals with various spatial resolutions.
Notably, our approach is applicable to other quantum sensors, such as spin defects in silicon carbide (SiC) \cite{Klimov2015,Widmann2015,Jiang2023} and hexagonal boron nitride (hBN) \cite{Gottscholl2020,Rizzato2023}, thus providing a benchmark for narrowband magnetic imaging based on dynamical decoupling with any spin sensor.

\section{Experimental Methods}
\begin{figure*}
    \centering
    \includegraphics[width=\linewidth]{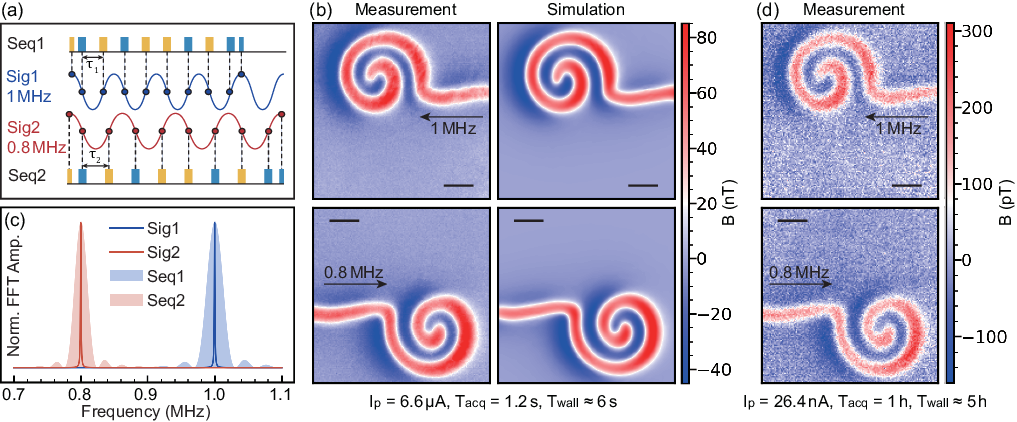}
    \caption{RF-QDM imaging of narrowband, multi-frequency magnetic fields of known phase, using a dynamical decoupling (XY8-8) protocol.
    (a) Simultaneously, a 1\,MHz alternating current (Sig1) is applied to the upper microcoil on the NV-diamond surface [Fig. \ref{fig:1}(a)]; and a 0.8\,MHz current (Sig2) is applied to the lower microcoil. 
    By adjusting the spacing between $\mathrm{\pi}$ pulses, the dynamical decoupling sequence becomes sensitive to different signal frequencies. 
    The diagram presented here illustrates simple XY8 sequences, while in the experiment, an XY8-8 sequence with phase modulation is employed, as depicted in [Fig. \ref{fig:1}(c)].
    (b) Comparison of measured (left) and simulated (right) magnetic field images resulting from alternating currents with a peak value of 6.6$\,\mathrm{\mu A}$.
    After averaging over 1.2\,s (for each signal frequency), the experimental images exhibit a SNR of approximately 50, calculated as the ratio of the maximum signal contrast to the standard deviation within a quarter of the FOV excluding the signal.
    By adjusting the $\pi$ pulse spacing $\tau$, the XY8-8 sensing sequence can be synchronized with either Sig1 or Sig2, enabling imaging of RF magnetic field patterns generated by each microcoil individually, with no observable ``cross-talk" in the images between the simultaneously applied signals.
    Due to the data transfer time from the camera to the computer, the wall time of this experiment is $\approx6\,$s.
    (c) Calculated FFT spectra of the two signals applied to the microcoils (blue and red solid lines); and filter functions of the XY8-8 sequences used for spectrally-selective RF-QDM measurements (shaded blue and red regions). Both signal and filter function amplitudes are normalized. A 10\,Hz linewidth is added to the signal peaks for better visualization. Here, we only show the first-order filter function; the XY8 sequences can also sense higher-order odd harmonics.
    (d) Demonstration of sub-nanotesla scale RF magnetic imaging. 
    By reducing the peak value of the applied alternating currents to 26.4$\,\mathrm{nA}$, images are obtained of the RF magnetic field patterns from Sig1 and Sig2, with SNR of about 15 after one hour of signal averaging (for each signal frequency).
    All scale bars correspond to 50$\,\mathrm{\mu m}$ in this figure. The colorbars in (b) and (d) are not centered around zero because of asymmetry in the projection of the RF signal magnetic field amplitude on the sensing NV axis.
    } 
    \label{fig:2}
\end{figure*}
The present study employs a custom-built QDM utilizing a CVD-grown diamond plate [Fig. \ref{fig:1}(a)] with a nitrogen-doped surface layer produced by Element Six, Ltd.
Unless otherwise noted, measurements and simulations in the main text are for a diamond with a $10\,\mathrm{\mu m}$ layer [$\mathrm{NV}\approx2.7\,$ppm].  
Results for a diamond with a $1.7\,\mathrm{\mu m}$ layer with the same $\mathrm{NV}$ concentration, as well as other RF-QDM technical details, are reported in the Appendix \ref{supp:thinNV}.
The negatively charged $\mathrm{NV^-}$ (otherwise referred to as NV throughout this paper) possesses a triplet ground-state spin [Fig. \ref{fig:1}(b)] that can be optically initialized and read out. In this work, 532\,nm green laser light is coupled into a polished side facet of the diamond plate, undergoing total internal reflection (TIR) and illuminating a $\sim{400\times600}\,\mathrm{\mu m^2}$ region of the NV-doped layer. 
A nominal 22\,mT bias magnetic field is aligned to one of the four NV ensemble orientations in diamond for sensing the projection of signal RF magnetic fields along that particular sensing NV orientation (defined as the z axis). 
The spin-state-dependent $\mathrm{NV}$ fluorescence is subsequently collected with an objective (20$\times$/0.75\,NA Nikon) and imaged with a lock-in camera (Heliotis heliCam C3) capable of external frame rates up to 3.8$\,$kHz.

An arbitrary waveform generator (AWG) is used to control the experimental apparatus, including direct synthesis of the microwave (MW) pulses that drive transitions between the NV $\mathrm{m}_\mathrm{s}=0$ and $\mathrm{m}_\mathrm{s}=+1$ sub-levels [Fig. \ref{fig:1}(b)]. 
The AWG marker channels synchronize the camera acquisition and gating of an acoustic optical modulator (AOM) used to generate optical pulses exciting the NV layer. 
In addition, the AWG produces independent RF test signals applied to two spiral-shaped microcoils on the NV-diamond surface, which provide spatially-varying magnetic field patterns across the FOV.

For each imaging pixel, the local RF magnetic fields generated by the microcoils can be measured using a dynamical decoupling protocol (XY8-n), where the net phase acquired by the $\mathrm{NV}$ electronic spin due to the oscillating field is mapped onto a difference in the ground state spin population by a final $\pi/2$ pulse prior to readout. 
Four different XY8-n sequences, comprising a single demodulation cycle, are interwoven between exposures of the lock-in camera, as depicted in [Fig. \ref{fig:1}(c)], and the corresponding NV fluorescence signals are labeled as $S_{1,2,3,4}$.
By controlling the phase of the final $\mathrm{\pi/2}$ pulse, one can enhance the readout signal contrast and subtract broadband laser intensity noise by leveraging the in-pixel lock-in camera demodulation. 
In each cycle and for every pixel, the camera generates an in-phase (I) signal (i.e., internal I frame) by analog
subtraction of the third exposure from the first; and a quadrature (Q) signal (i.e., internal Q frame) as the difference of the second and fourth exposures.
The accumulated difference signals after $N_d$ demodulation cycles are digitized to produce external I and Q frames, which are stored in an on-board, memory buffer. 
Finally, since each I and Q frame pair contains the same information due to the specific phase alternation pattern employed, they are summed to form one external frame of the readout signal after transfer to the host computer.
Thus, there are $2N_d$ total internal frames for each external frame.
For an applied RF test signal with $B_\mathrm{RF} = B_z\cos(2\pi f_st + \phi_0)$, where $B_z$ is the projection of the signal magnetic field amplitude on the NV axis, $f_s$ is the oscillation frequency, and $\phi_0$ is the phase offset, the readout signal for each pixel of an external frame can be written as $S= N_d[(S_1-S_3)-(S_2-S_4)]=4N_dS_a\sin(\kappa B_z\cos\phi_0)$, where $S_a$ is the signal oscillation amplitude, $\kappa = 2\gamma N_\mathrm{\pi}/f_s$ is a constant, $\gamma = 2\pi\times 28.024$\,GHz/T \cite{Acosta2010} is the NV electronic gyromagnetic ratio, and $N_\mathrm{\pi}$ is the $\mathrm{\pi}$ pulse number in the XY8-n sequence.
When sensing a small magnetic field shift $\Delta B_z$, the change in NV fluorescence readout is given by $\Delta S = 4N_d S_a \kappa \Delta B_z$. 
By calibrating the magnetometry slope of each pixel, the RF magnetic field image can be extracted from the fluorescence image as shown in the Appendix \ref{supp:DDAC}.

The CASR protocol \cite{Glenn2018} is employed for simultaneous near-field imaging of the amplitude, frequency, and phase of RF signal patterns, with high spectral resolution ($\approx1$\,Hz). 
By recording consecutive lock-in camera external frames [Fig. \ref{fig:1}(c)], the time-dependent RF magnetic field is captured by interspersed NV spin-state readouts at a sampling rate of $f_\mathrm{SR}$, equal to the external frame rate. 
Using the CASR protocol, an oscillating NV fluorescence signal in the time domain is read out for each pixel. 
Applying a Fast Fourier Transform (FFT) yields frequency domain CASR spectra for each pixel, centered at the alias frequency $f_s - mf_\mathrm{SR}$, with peak amplitude $S_{\mathrm{CASR}}=4N_d(2\pi)^{\frac{3}{2}}S_a J_1(\kappa B_z)$ (see Appendix \ref{supp:CASR} for details), where $J_1(x)$ is the first-order Bessel function and $m$ is the integer closest to $f_s/f_\mathrm{SR}$. 
For the demonstrations presented here, we experimentally ensure that $\kappa B_z < 1$ such that the first-order Bessel function closely approximates a linear function, leading to $S_{\mathrm{CASR}} \propto B_z$.

\section{Results}
\subsection{In-phase RF magnetic imaging}
\label{RFImaging}

To characterize the performance of the RF-QDM, we begin by imaging narrowband magnetic fields with known phase. 
We simultaneously apply alternating current (AC) signals at $\mathrm{1\,MHz}$ and $\mathrm{0.8\,MHz}$, each with a peak value of $\mathrm{6.6\,\mu A}$, to two separate microcoils on the NV-diamond surface [Fig. \ref{fig:2}(a)]. 
We first align the XY8-8 sensing sequence in-phase with the $\mathrm{1\,MHz}$ signal by setting the $\mathrm{\pi}$ pulse spacing to $\tau_1 = 0.5\,\mathrm{\mu s}$ and sweep the test signal phase to obtain maximum signal.
After a 1.2 second acquisition (200 camera external frames), we obtain an RF magnetic field amplitude image with a peak-to-peak difference of about 130$\,$nT, revealing the near-field pattern generated by \textit{only} the upper microcoil with $\mathrm{SNR}\approx{50}$ [Fig. \ref{fig:2}(b)]. 
Adjusting the spacing between XY8-8 $\mathrm{\pi}$ pulses ($\tau_2 = 0.625\,\mathrm{\mu s}$) to be in phase with the $\mathrm{0.8\,MHz}$ signal, we record the magnetic field pattern produced \textit{only} by the lower microcoil. The RF-QDM provides clean near-field images of either signal frequency, with no significant ``cross-talk" between the simultaneously-applied signal patterns, because the test signal frequencies are spectrally separated by more than the narrowband filter function of the sensing sequence [Fig. \ref{fig:2}(c)]. 
The center frequency of the filter function is easily tunable, as it is determined by the spacing $\tau$ between pulses in the sensing sequence; while the bandwidth (spectral detection window) is determined by the number of repetitions, $n$, of the XY8-n sequence \cite{Lukasz2008}.
 For the present demonstration, using an XY8-8 sensing sequence and signal frequencies $\sim{1}$\,MHz, the filter function bandwidth $\approx30$\,kHz. 
This demonstration illustrates the ability of the RF-QDM to perform micron-scale spatial discrimination of spectrally well-separated narrowband signals. 

In [Fig. \ref{fig:2}(b)], the measured RF magnetic field distributions are compared to numerical simulations. 
We calculate the three-dimensional magnetic field  generated by a static current of $\mathrm{6.6\,\mu A}$ constrained to the physical geometry of the spiral-shaped microcoils using finite element software \cite{comsol}. 
Averaging the magnetic field at a $\mathrm{10\,\mu m}$ stand-off distance over the $\mathrm{10\,\mu m}$ NV layer thickness and projecting the field vector onto the sensing NV orientation, the simulated magnetic field images are consistent with measurements. 
To highlight the sensitivity and stability of the RF-QDM, we reduce the amplitude of the two test signals to 26.4$\,$nA. 
Following an acquisition time of 3600$\,$s for each signal frequency, we obtain images of the in-phase narrowband  magnetic field amplitude generated by each microcoil, with a peak-to-peak difference of about 500$\,$pT and $\mathrm{SNR}\approx{15}$ [Fig. \ref{fig:2}(d)]. 
Note that a machine learning approach has recently been applied to the inverse problem of reconstructing the current distribution from QDM magnetic field images \cite{reed2024}. 

\subsection{RF magnetic imaging performance}
\begin{figure}
    \centering
    \includegraphics[width=\linewidth]{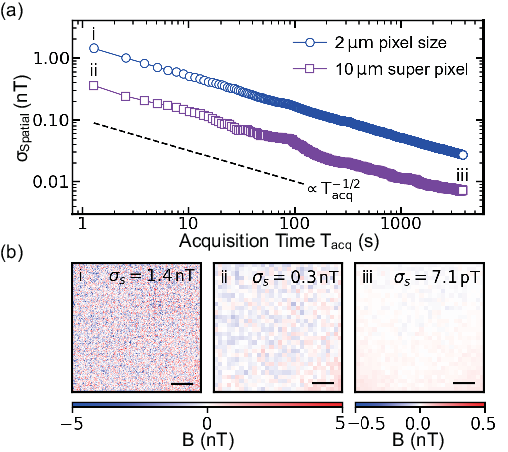}
    \caption{
    Suppressing spatial noise in RF-QDM magnetic images by averaging over time and/or space.
    (a) Measured magnetic spatial noise floor $\sigma_\mathrm{Spatial}$ with no RF signal applied to the microcoils and a 1\,MHz detection frequency, as a function of data acquisition time $T_\mathrm{acq}$, for $\mathrm{2\,\mu m}$ single pixel resolution and $\mathrm{10\,\mu m}$ (binned) super pixel resolution.
    A dashed black line depicts the power law scaling behavior $\propto T_\mathrm{acq}^{-1/2}$ as a guide to the eye.
    (b) Example RF-QDM magnetic images of the noise background and resulting $\sigma_\mathrm{Spatial}$: i) 2$\,\mathrm{\mu m}$ single pixel resolution with $T_\mathrm{acq} = 1.2\,\mathrm{s}$; ii) 10$\,\mathrm{\mu m}$ super pixel resolution with $T_\mathrm{acq} = 1.2\,\mathrm{s}$; iii) 10$\,\mathrm{\mu m}$ super pixel resolution with $T_\mathrm{acq} = 3600\,\mathrm{s}$.
    Corresponding values for $\sigma_\mathrm{Spatial}$ are labeled in (a).
    All scale bars correspond to 50$\,\mathrm{\mu m}$ in this image.
    The magnetic noise is distributed randomly without significant spatial correlation.
    The comparison between i) and ii) shows that spatial noise can be averaged over space, while the comparison between ii) and iii) illustrates an average over time.
    Note that the frame rate of the lock-in camera is constant in all measurements}
    \label{fig:3}
\end{figure}
\begin{figure*}
    \centering
    \includegraphics[width=\linewidth]{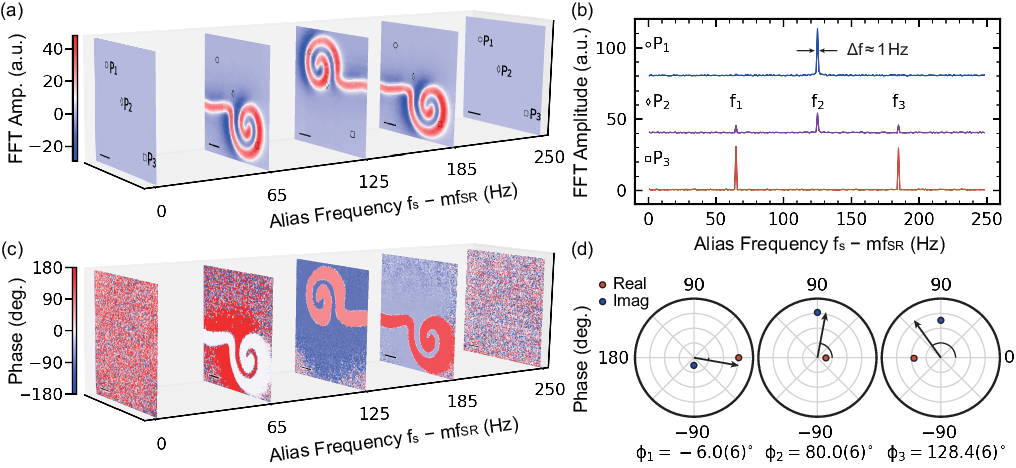}
    \caption{RF-QDM imaging of arbitrary phase, multi-frequency magnetic fields using the CASR protocol.
    As described in the main text, a single tone alternating current (1\,MHz + 125\,Hz) is applied to the upper microcoil; and a two tone current (1\,MHz + 65\,Hz and 1\,MHz + 185\,Hz) is simultaneously applied to the lower microcoil. 
    Both currents have peak amplitude $I_p = \mathrm{3.3\,\mu A}$.  
    The phases of each signal frequency component (tone) are arbitrary but fixed during the measurements.
    Total image acquisition time is 200\,s.
    (a) CASR images of RF magnetic field amplitudes at each signal frequency component, as well as at alias frequencies  (1\,Hz and 250\,Hz) far from any signal tone. 
    The imaged arbitrary phase magnetic field amplitudes have SNR of approximately 70 (calculated with the same method as [Fig. \ref{fig:2}(b)]) for each signal frequency component, and exhibit good agreement with images of in-phase signals shown in [Fig. \ref{fig:2}(b)]. The colorbar is not centered around zero because of asymmetry in the projection of the RF magnetic field amplitude on the sensing NV axis.
    (b) CASR spectra (FFT) of three test pixels $\mathrm{P_{1,2,3}}$ with positions labeled in (a). 
     Peaks corresponding to three frequency components $f_{1,2,3}$ are clearly illustrated in the spectra, with peak heights proportional to the RF magnetic field amplitude. 
     Linewidths of all spectral peaks $\approx1$\,Hz. 
     Amplitude offsets are added to the spectra of $\mathrm{P_{1,2}}$ for a better visualization.
    (c) CASR images of RF magnetic field phases at each signal frequency component, as well as at alias frequencies far from any signal tone (1\,Hz and 250\,Hz).
    Noisy phase distributions in no-signal regions are attributed to the ratio of near-zero real and imaginary amplitudes of the CASR FFT.
    (d) Polar plots illustrate the phases of each signal frequency component, derived from CASR measurements. 
    $\phi_{1}$ and $\phi_{3}$ are determined from the real and imaginary parts of the FFT of pixel $\mathrm{P_3}$; and $\phi_{2}$ is determined from $\mathrm{P_1}$. 
    Red and blue dots represent the real and imaginary amplitudes of the associated FFT spectra, respectively, while black arrows indicate the absolute value of the amplitude as shown in (a). 
    Grids are depicted as circles with a step size equivalent to the radius of 10 arbitrary units of FFT amplitude.
    All scale bars in (a) and (c) correspond to 50$\,\mathrm{\mu m}$.  
    } 
    \label{fig:4}
\end{figure*}

With a $20\times$ magnification objective and $39.6\,\mathrm{\mu m}$ camera pixel size, the RF-QDM has a $2\,\mathrm{\mu m}$ lateral optical resolution, corresponding to the size of a single pixel in the magnetic images [Fig. \ref{fig:2}(b,d)].
For the diamond with a $10\,\mathrm{\mu m}$ NV layer diamond, each sensing voxel is an approximate $2\times2\times10\,\mathrm{\mu m^3}$ region; in this case, a single camera pixel can collect fluorescence from adjacent NV sensing voxels, resulting in QDM-RF spatial resolution greater than $2\,\mathrm{\mu m}$.  
For the $1.7\,\mathrm{\mu m}$ NV layer diamond, QDM-RF spatial resolution $\approx2\,\mathrm{\mu m}$ (see Appendix \ref{supp:thinNV}). 
For the dynamical-decoupling-based experiments discussed above, we vary the number of XY8 sequences and measure the narrowband sensitivity of each pixel across the FOV for a 1\,MHz signal frequency, as shown in the Appendix \ref{supp:ACSensOpt}. 
The optimized sequence is determined to be XY8-8, with a median per-pixel sensitivity of $0.93(1)\,$nT$\cdot$Hz$^{-1/2}$.
Given a sensing voxel volume $\approx40\,\mathrm{\mu m}^3$ for the $10\,\mathrm{\mu m}$ NV layer diamond, the volume-normalized sensitivity is about 6\,nT$\cdot$Hz$^{-1/2}\cdot\mathrm{\mu m}^{3/2}$, which is comparable to state-of-the-art ensemble NV sensitivity using XY8 sequences and photodiode (non-imaging) measurements \cite{Arunkumar2023}. 
We estimate the contribution of two per-pixel noise sources: photon shot noise $\approx0.3\,$nT$\cdot$Hz$^{-1/2}$ and camera quantization noise $\approx0.86\,$nT$\cdot$Hz$^{-1/2}$, indicating that the present RF-QDM's narrowband sensitivity is limited by the camera.
A similar performance analysis is presented in the Appendix \ref{supp:ACSensOpt} for the RF-QDM using the 1.7\,$\mathrm{\mu m}$ NV layer diamond. 

In addition to per-pixel sensitivity, spatial noise (due to pixel-to-pixel variance from the camera and spatially-varying diamond crystal stress) is another significant parameter for evaluating RF-QDM performance. 
To characterize the spatial noise, we conduct single-frame measurements without an applied RF magnetic signal. 
At a given narrowband detection frequency, the spatial noise floor of an RF-QDM image, $\sigma_\mathrm{Spatial}$, is defined as the pixel-to-pixel standard deviation of the measured in-phase magnetic field amplitude at the detection frequency, determined across the entire FOV; and is inversely proportional to the square root of the averaging frame number.
During magnetic imaging, a trade-off arises between spatial resolution and spatial noise. 
One strategy to mitigate spatial noise is to bin pixels: i.e., group adjacent pixels together to create ``super pixels"; and thereby, effectively reduce the overall spatial noise in the image, albeit with coarser spatial resolution. 
For applications where the highest spatial resolution of the RF-QDM is not required, averaging over the space domain rather than the time domain allows for shorter acquisition times with similar SNR.
We demonstrate this approach for 5$\times$5 binned super pixels of size 10$\,\mathrm{\mu m}$.
With no signal applied to the microcoils, we observe $\sigma_\mathrm{Spatial} \propto T_\mathrm{acq}^{-1/2}$ for RF-QDM images and a 1\,MHz detection frequency, where $T_\mathrm{acq}$ is the data acquisition time for both 2\,$\mathrm{\mu m}$ (unbinned) pixels and 10\,$\mathrm{\mu m}$ super pixels [Fig. \ref{fig:3}(a)].
This behavior indicates that the RF-QDM's spatial magnetic noise can be suppressed with time averaging out to hour timescales.
The spatial noise floor of a single frame with 1.2$\,$s acquisition time and 10$\,\mathrm{\mu m}$ super pixel resolution is similar to that of a 30$\,$s average of multiple frames with 2$\,\mathrm{\mu m}$ single pixel resolution.
With 10$\,\mathrm{\mu m}$ super pixel spatial resolution and 1\,hour acquisition time, the RF-QDM can average the spatial noise background down to 7.1(1)\,pT [Fig. \ref{fig:3}(b)].
The median per-pixel narrowband sensitivity for a 10$\,\mathrm{\mu m}$ super pixel is also improved to $0.18(1)\,$nT$\cdot$Hz$^{-1/2}$.

\subsection{Simultaneous amplitude and phase imaging}
\label{CASRImaging}
Next, we implement the CASR protocol in the RF-QDM, and simultaneously image the amplitude and phase of a multi-frequency RF magnetic field, with spectral resolution $\approx1$\,Hz as shown in [Fig. \ref{fig:4}].
For these demonstration measurements, we simultaneously apply a dual-tone alternating current $I_1 = I_p\cos[2\pi(f_0+f_1)t+\theta_1]+I_p\cos[2\pi(f_0+f_3)t+\theta_3]$ to the lower microcoil, and a single-tone current $I_2 = I_p\cos[2\pi(f_0+f_2)t+\theta_2]$ to the upper microcoil. 
Here, $f_0 = 1\,$MHz is the carrier frequency; $f_{1,2,3}$ are 65$\,$Hz, 125$\,$Hz and 185$\,$Hz, respectively; and $\theta_{1,2,3}$ are arbitrary phase offsets of each frequency component, which are fixed during the measurements.
The current peak value of $I_p = \mathrm{3.3\,\mu A}$ is half that of the in-phase signals used in the above results [Fig. \ref{fig:2}(b)], with estimated peak RF magnetic field amplitude less than 40$\,$nT, which is in the linear response range of the measurement.
The lock-in camera frame rate is set to $f_{\mathrm{SR}}=500$\,Hz with 8 demodulation cycles per frame.
After 1 second of measurement, a series of NV fluorescence images are acquired [Fig. \ref{fig:1}(c)], which gives the RF magnetic field change in the time domain, sampled at the CASR alias frequency.
We repeat the measurement 100 times to average down the noise floor, then perform an FFT to the measured time-dependent fluorescence change of each pixel.
Finally, we normalize the CASR signal amplitude to the steady-state NV fluorescence for each pixel (in the absence of any NV spin-state modulation) to extract the magnetic field amplitude images of each frequency component, as described in the Appendix \ref{supp:CASR}.

Three test pixels are selected to illustrate the CASR amplitude imaging results [Fig. \ref{fig:4}(a,b)]. 
For pixel $\mathrm{P_1}$, which is located at the upper microcoil, only one frequency peak at $f_2$ is observed in the CASR spectrum.
Pixel $\mathrm{P_3}$, located at the lower microcoil, exhibits two frequency components in its CASR spectrum, corresponding to $f_1$ and $f_3$. 
The central pixel $\mathrm{P_2}$, located between the two microcoils, exhibits frequency components from both sources, with reduced amplitude.
In all cases, the linewidth of CASR spectral peaks $\Delta f\approx1$\,Hz. 
From the CASR spectra of each pixel, we can generate magnetic field amplitude images for each frequency component within the measurement bandwidth, as shown in [Fig. \ref{fig:4}(a)] for five different example frequencies. 
The illustration of CASR amplitude images at frequencies $f_{1,2,3}$ provides insight into the RF magnetic field patterns generated by both microcoils, consistent with the in-phase magnetic field images and simulation results presented in [Fig. \ref{fig:2}(b)].

The CASR protocol also enables simultaneous magnetic field phase imaging for each frequency component within the measurement bandwidth. 
The phases of each measured frequency component, $\phi_{1,2,3}$, are dependent on the phases, $\theta_{1,2,3}$, of the input test signals, the impedance of the microcoils, and the time delay between the current signal and the CASR sensing sequence. 
All of these parameters are fixed during measurement, which provides arbitrary but stable phases for the imaged RF magnetic field patterns.
When extracting the CASR amplitude from time-domain NV fluorescence measurements, it is essential to consider the phase-dependent real and imaginary components of the CASR FFT spectra.
The root mean square of these components yields the amplitude, while the phase can be extracted from the ratio. 
Based on the measured real and imaginary components of the CASR spectra at locations P$_1$ and P$_3$, we determine phases $\phi_{1,2,3}$ for each signal frequency components with accuracy of $0.6^\circ$ calculated from the noise floor in [Fig. \ref{fig:4}(b)].
Repeating the same phase calculation for each pixel across the FOV, we obtain phase images at each frequency step of the CASR spectra, as shown in [Fig. \ref{fig:4}(c)].
Further details are discussed in the Appendix \ref{supp:CASRDetails}.

\section{Discussion}
\begin{figure}
    \centering
    \includegraphics[width=\linewidth]{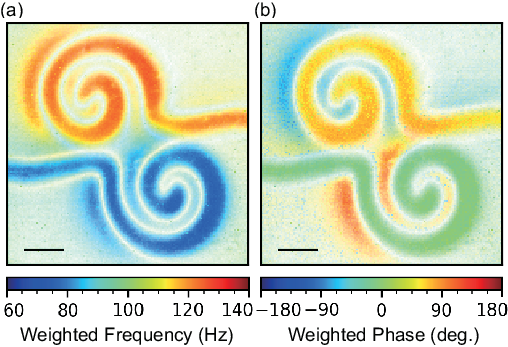}
    \caption{RF-QDM CASR imaging of weighted frequency and phase of magnetic signals generated by the two microcoils for the same experimental conditions and CASR imaging protocol as in [Fig. \ref{fig:4}].
    (a) The weighted frequency $f_w = \int f\cdot S_\mathrm{CASR}\,df/\int S_\mathrm{CASR}\,df$.
    A transparency mask, based on the integral of the CASR amplitude of each pixel, is applied to the image data to filter out the no-signal region for better visualization. 
    The colorbar indicates the frequency integration range, which in this example includes $f_1$ and $f_2$, but excludes $f_3$. 
    The near-field, micron-scale spatial distribution of the two included frequency components is clearly illustrated in the image.
    (b) Weighted phase image of the RF magnetic signal generated by the two microcoils. 
    The weighted phase $\phi_w = \int \phi \cdot S_\mathrm{CASR}\,df/\int S_\mathrm{CASR}\,df$.
    The transparency mask and integration range are the same as in (a). 
    The spatial distribution of $\phi_1=-6^\circ$ and $\phi_2=80^\circ$ is clearly illustrated.
    The $\sim{180^\circ}$ phase shifted signal outside each spiral is due to the opposite direction of RF magnetic field generated by the microcoils.
    All scale bars correspond to 50$\,\mathrm{\mu m}$ in this image.
    } 
    \label{fig:5}
\end{figure}
A promising application of the RF-QDM is micron-scale real-space NMR imaging.  
 The RF-QDM's Hz-scale spectral resolution will enable micron-scale spatial identification of chemical shifts and J-couplings in samples on the diamond surface \cite{Glenn2018} — key spectral identifiers of molecular structure that are critical to NMR applications in chemistry, life science, and materials research.
Based on the results from previous NV-NMR research \cite{Glenn2018,Bucher2020}, we estimate a spatial noise threshold of 50$\,$pT for resolvable micron-scale NMR images with $\mathrm{SNR}>1$ for a thermally polarized sample in an external (bias) magnetic field $\sim{0.1}$\,T; and a spatial noise threshold of 10 nT if using dynamic nuclear polarization (DNP) of the sample.
Assuming the current RF-QDM performance, an acquisition time around 40\,s is required for a thermally polarized sample to surpass this threshold with spatial resolution of 10$\,\mathrm{\mu m}$; whereas DNP hyperpolarization would allow the threshold to be passed in about 30\,ms with 2$\,\mathrm{\mu m}$ spatial resolution [Fig. \ref{fig:3}(a)].
(Other hyperpolarization techniques can provide even greater nuclear spin polarization, and NV-NMR signal amplitude \cite{Arunkumar2021}, albeit with operation typically limited to lower bias fields.)
With the millisecond-scale time resolution of the RF-QDM's lock-in camera, it may thus be possible to image real-time NMR dynamics in micron-scale samples, including chemical processes in living systems at the cellular scale and samples within microfluidic structures \cite{Briegel2024}.
To demonstrate the ability of the RF-QDM to spatially discriminate narrowband signals that are spectrally crowded (i.e., differ in frequency by a few tens of Hz, as in many NMR signals of interest), we show example CASR images of weighted frequency and phase from the two microcoils with 60 Hz offset in their signals (around a 1\,MHz carrier frequency) [Fig. \ref{fig:5}].
The near-field RF signals produced by the two microcoils are clearly distinguished at the micron scale by their different frequencies and phases. 
Similar weighted frequency (or phase) NV-NMR images could allow detailed spatial monitoring of chemical concentratons in samples of interest.

In this work, the amplitude of RF magnetic signals is held constant during measurements.
However, it is worth noting that RF-QDM measurement protocols are applicable to signals with decay envelopes, such as free-inductive-decay (FID) in NMR experiments.
A time domain decay envelope manifests as finite FFT peak linewidths in the CASR spectrum of each pixel, yielding spatial information from both the imaged signal frequencies and linewidths. 
This capability of real-space imaged NV-NMR spectroscopy holds promise for applications involving micron-scale samples with spatially varying concentrations and local environments of NMR-detectable nuclear species.
Furthermore, technical advances in lock-in camera frame rate and storage capacity can lead to corresponding improvements in sensing bandwidth and spectral resolution, for micron-scale NMR and other applications.  
In addition, the RF-QDM should be compatible with magnetic field gradients applied to the NV spins \cite{Arai2015,Zhang2017}, potentially allowing combined Fourier and real-space NMR imaging with nanoscale spatial resolution and FOV approaching a millimeter.

We also highlight other promising applications of the RF-QDM.
Previous work demonstrated single-channel measurement of the AC susceptibility of 2D magnetic materials using NV centers and XY8 sequences \cite{PRXQuantum.2.030352}. 
The RF-QDM will allow micron-scale mapping of the spatial distribution of AC susceptibility, which could serve as an important calibration process for material science investigations.
The RF-QDM can also be applied to imaging inductive eddy currents in electrical conductors \cite{WangC2018}. 
By simultaneously imaging the amplitude and phase of the inductive AC magnetic signal, the RF-QDM could provide higher spatial resolution and better sensitivity than conventional techniques for eddy current inspection.
For integrated circuits and other micron-scale electronics, the RF-QDM may be useful for failure analysis \cite{Oliver2021,PhysRevApplied.20.014036} and hardware Trojan detection \cite{Ashok2022}.
Mapping circuit magnetic field activity across various operating frequencies, the RF-QDM could provide complementary information to static field images \cite{Turner2020,Kehayias2022} and other electromagnetic properties.
Additionally, the RF-QDM can be applied to impedance tomography, which is widely used in biological science \cite{Bounik2022}. 
Instead of using arrays of coplanar electrodes to detect the spatial distribution of modulated stimulation currents in living cells and tissue — which is invasive and only provides a limited set of point measurements — the RF-QDM may provide a noninvasive direct image of the current amplitude and phase, with full spatial coverage across the FOV.

In summary, we characterize the performance of a lock-in-camera-based quantum diamond microscope (QDM) for narrowband magnetic field imaging; and demonstrate imaging protocols for multi-frequency widefield imaging of near-field patterns of radiofrequency (RF) signals generated by fabricated microcoils, with $\approx1$\,Hz spectral resolution. 
This RF-QDM operates under ambient conditions and provides per-pixel sensitivity to narrowband fields $\sim{1}\,$nT$\cdot$Hz$^{-1/2}$ with $\approx2$\,$\mathrm{\mu m}$ spatial resolution across a 300$\times$300$\,\mathrm{\mu m^2}$ field-of-view (FOV).
With temporal averaging and spatial binning, the spatial noise floor is reduced to the picotesla-scale, similar to the expected signals from potential applications of real-space NV-NMR imaging using thermally-polarized samples.
The RF-QDM simultaneously captures the micron-scale spatial distribution of the amplitude, frequency, and phase of RF magnetic fields, also paving the way for other application areas such as AC susceptibility imaging, eddy current inspection, assessment of electronic circuits, and impedance tomography.

Note that for the results presented here, the RF magnetic field amplitude images depict the projected value on the sensing NV orientation. 
However, as there are four NV orientations in diamond, the RF-QDM should be applicable to imaging of the RF magnetic field vector \cite{Schloss2018}.
This capability, to be demonstrated in future work, could also allow the correction of image artifacts related to variation of RF magnetic field direction across the finite NV layer.
In addition, the present RF-QDM is limited to signal frequencies $<10$\,MHz by technical constraints on the NV Rabi frequency.
Future advances in delivered MW signal strength (and hence increased NV Rabi frequency), or the application of a quantum frequency mixing protocol \cite{WangG2022}, will allow imaging of higher signal frequencies ($\sim{10}$\,MHz to GHz range).
We also note that the RF-QDM is compatible with narrowband sensing protocols that provide improved sensitivity compared to XY8, e.g., DROID-60 \cite{Zhou2020,Choi2020} and quantum logic enhanced repetitive readout \cite{Arunkumar2023}.
However, these protocols typically have increased complexity, resulting in a longer sensing time.
Currently, due to limitations imposed by the memory of the RF-QDM's lock-in camera, employing such protocols would reduce the frame rate and available measurement time, leading to a reduced spectral detection window for CASR imaging.
Improvements in camera technology and data processing algorithms may help alleviate these limitations, further enhancing RF-QDM performance for narrowband magnetic field imaging.

\begin{acknowledgments}
We acknowledge the Maryland NanoCenter and its FabLab for providing instruments and assistance to fabricate the microwave waveguide and microcoils. This work is supported by, or in part by, the U.S. Army Research Laboratory under Contract No.  W911NF1920181; the U.S. Army Research Office under Grant No. W911NF2120110; the U.S. Air Force Office of Scientific Research under Grant No. FA9550-22-1-0312; the Gordon \& Betty Moore Foundation under Grant No. 7797.01; and the University of Maryland Quantum Technology Center.
\end{acknowledgments}
\appendix

\section{Experimental Details} 
\label{supp:exp}

Two side-polished diamond samples are employed in this work: one contains a $10\,\mathrm{\mu m}$-thick, $^{15}$N-enriched CVD layer of NV centers ($[\text{N}]$~$=$~17\,ppm, $>$\,99.99\% $^{12}$C), grown by Element Six Ltd.~on a $2\times2\times0.5$\,mm$^3$ high-purity diamond substrate, with post-growth treatment via electron irradiation and annealing to increase the NV concentration to 2.7\,ppm;
the other diamond sample contains a $1.7\,\mathrm{\mu m}$-thick NV layer with other properties being the same.

A diode-pumped solid-state laser (Lighthouse Photonics, Sprout-H) is employed to generate a continuous-wave 532\,nm laser beam. 
The laser beam is focused onto an acousto-optic modulator (AOM) (Gooch \& Housego, Model 3250-220) by a $f=500$\,mm spherical convex lens; and the first-order diffracted beam from the AOM is used for the experiments. 
To create optical pulses for NV initialization and readout, the radio frequency (RF) driver for the AOM is gated by a switch (Mini-Circuits, ZASWA-2-50DR+) that is controlled by transistor-transistor logic (TTL) pulses from the marker channel of an arbitrary waveform generator (AWG) (Zurich Instruments, SHFSG). 
The first-order beam after the AOM is further shaped by an $f=130$\,mm cylindrical convex lens and a $f=100$\,mm spherical convex lens to produce an elliptical beam profile incident on a polished side surface of the diamond sample. 
Approximately 2.5\,W of laser light enters the diamond and undergoes total internal reflection at the surface with the NV layer, illuminating an area of $\sim{400\times600}\,\mathrm{\mu m^2}$. 
NV fluorescence (also commonly referred to as photoluminescence or PL) is collected by a 20$\times$/0.75\,NA Nikon objective, and filtered by a 647\,nm long-pass filter (Semrock) before projecting onto a photodiode (Thorlabs, PDA36A2) or a lock-in camera (Heliotis, heliCam C3). 
Photodiode voltage measurements are acquired using a DAQ system (National Instruments, NI USB6259). 
 
\begin{figure}
    \centering
    \includegraphics[width=\linewidth]{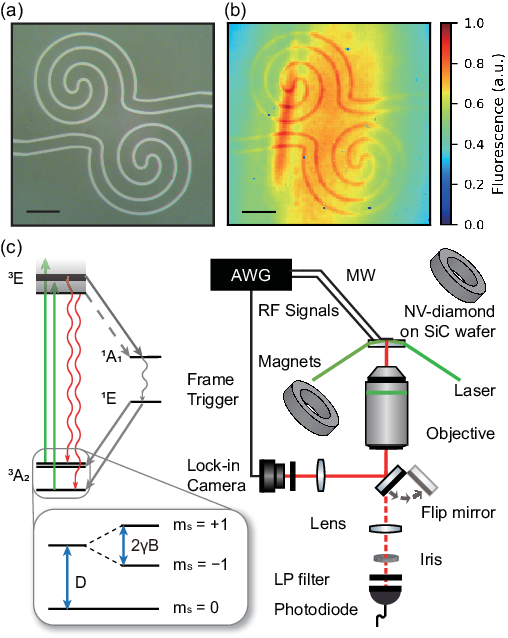}
    \caption{
    (a) Optical image of the spiral-shaped microcoils with a stand-off distance $\approx10\,\mathrm{\mu m}$  to the NV layer. All scale bars correspond to 50 $\mathrm{\mu}$m.
    (b) NV fluorescence image under continuous wave (CW) laser excitation illustrating the relative illumination of the NV layer (i.e., beam shape) in the field-of-view (FOV). The normalized fluorescence is proportional to the photon count.
    (c) NV energy-level diagram and room temperature experimental apparatus. 
    The NV center in diamond exhibits spin-dependent fluorescence with triplet ground-state spin transitions. 
    A flip mirror directs NV fluorescence towards either a photodiode or a lock-in camera. 
    An AWG generates MW pulses to control NV spin, RF signals for the microcoils, and frame triggers for camera exposure, utilizing its internal oscillator to synchronize the experiment. A SiC wafer serves as a mount for the MW loop, microcoils, and NV-diamond chip. An RF test coil (not shown) provides signals for calibration of AC magnetometry.}
    \label{fig:6}
\end{figure}

Microwave (MW) pulses for NV spin control are generated by a signal channel of the same AWG used for generating TTL pulses. A fabricated 300\,nm-thick, $800\,\mathrm{\mu m}$ inner diameter gold \textOmega-shaped coplanar waveguide delivers MW pulses to control the NV electron spins, generating a homogeneous MW driving (Rabi frequency) region covering the FOV.
The fabricated palladium spiral microcoils, used for generating spatial patterns of radiofrequency (RF) signals, have a trace width of $5\,\mathrm{\mu m}$, span a $280\times300\,\mathrm{\mu m^2}$ area, and have a thickness of 10\,nm [Fig. \ref{fig:6}(a)].
Palladium is used because of its relatively high resistance (e.g., compared to gold) in order to avoid large mutual inductance between the microcoils and the MW coil, which could degrade performance of the MW coil.
The microcoils and MW loop are fabricated on a SiC wafer for mechanical support and heat dissipation. 
The diamond is glued to the surface of the MW loop with an estimated stand-off distance of $10\,\mathrm{\mu m}$. 
To generate RF magnetic field patterns, multi-frequency voltages produced by other signal channels of the AWG are applied to the microcoils to induce currents. 
An RF test coil, wound from 22 American Wire Gauge with 14 loops and a diameter of 50$\,$mm, is placed on the top of the SiC wafer to generate test signals for magnetometry curve calibrations as shown in Appendix \ref{supp:DDAC}.
A bias magnetic field $\approx22\,$mT is provided by four pairs of temperature-compensated, permanent samarium cobalt ring magnets. 
Ring magnet pairs provide a more homogeneous bias magnetic field than cylinder-shaped magnets.

The size of a single camera pixel in the objective focal plane is determined using the known size of the microcoils and an optical image visualized using NV fluorescence.
The metallic surfaces of microcoil traces reflect the excitation laser and increase the local illumination intensity, resulting in greater NV fluorescence with a pattern following the phantom structure as shown in [Fig. \ref{fig:6}(b)].
The area occupied by each camera pixel corresponding to the focal plane is $2\times2\,\mathrm{\mu m^2}$, which agrees with the estimate using the magnification of the objective ($20\times$) and the physical size of each pixel ($39.6\times39.6\,\mathrm{\mu m^2}$) as reported by the camera manufacturer.
A schematic of the experimental apparatus is shown in [Fig. \ref{fig:6}(c)].

\section{Dynamical-decoupling-based AC Magnetometry} 
\label{supp:DDAC}
\begin{figure*}
    \centering
    \includegraphics[width=\linewidth]{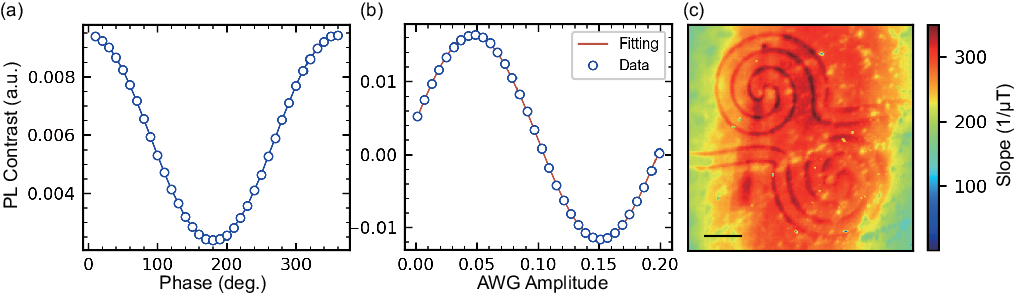}
    \caption{Example calibration measurements with the RF-QDM for dynamical-decoupling-based AC magnetometry (photodiode measurements with a laser beam illumination diameter $\approx50\,\mathrm{\mu m}$) and imaging (lock-in camera measurements). 
    (a) Phase sweep of photodiode AC magnetometry to determine the initial phase of a nominally in-phase AC magnetic field ($=1$\,MHz), applied with the RF test coil.
    (b) Photodiode AC magnetometry curve of relative NV photoluminescence contrast using the XY8-8 sequence, showing a good agreement with a sinusoidal fit.    
    (c) Imaged spatial distribution of the zero-crossing slope of the AC magnetometry curve, which exhibits a spatial dependence due to the laser beam illumination of the NV layer, see [Fig. \ref{fig:6}(b)]. The scale bar corresponds to 50$\,\mathrm{\mu m}$.
    }
    \label{fig:7}
\end{figure*}
The negatively charged nitrogen vacancy defect of interest (referred to simply as NV throughout this paper) has electronic spin S = 1, and its triplet ground state exhibits a zero-field splitting between the $\ket{m_s = 0}$ and $\ket{m_s = \pm 1}$ magnetic sublevels [Fig. \ref{fig:6}(c)]. The application of a bias magnetic field further splits the degenerate $\ket{m_s = \pm 1}$ states via the Zeeman effect. We use a pair of ring magnets to apply a bias magnetic field $\approx22\,$mT aligned along one of the four NV orientations in the diamond crystal. In this bias magnetic field regime, the effects of transverse diamond crystal strain and electric field can be neglected \cite{Kehayias2019,Barfuss2019}, such that the approximate Hamiltonian for the NV ground state is given by \cite{Tang2023,Doherty2013,Hart2021}:
\begin{equation}
\label{eq:NVHamiltonian}
\begin{aligned}
\frac{\hat{H}}{\hbar} \approx  (D+M_z) \hat{S}_z^2+\frac{\gamma}{2\pi} B_z\hat{S_z},
\end{aligned}
\end{equation}
where $\hat{S_z}$ is the dimensionless spin-1 operator along the NV quantization axis, $\gamma = 2\pi\times 28.024$\,GHz/T \cite{Acosta2010} is the NV electronic gyromagnetic ratio, $D\mathrm{\approx 2.87\,GHz}$ is the zero-field splitting at room temperature \cite{Acosta2010}, $B_z$ is the bias magnetic field, and $M_z$ is the longitudinal component of the crystal stress.

The dynamical-decoupling-based protocol for AC magnetometry begins with an optical pulse to initialize the electronic spin state to $\ket{m_s = 0}$ for each NV in the ensemble being probed. 
Subsequently, a $\mathrm{\pi/2}$ MW pulse prepares each NV in the ensemble in a superposition between $\ket{m_s = 0}$ and $\ket{m_s = 1}$ for AC magnetic field sensing.
The AC signal to be detected is an oscillating magnetic field $B_\mathrm{AC} = B_z\cos(2\pi f_st + \phi_0)$, where $B_z$ is the projection of magnetic field amplitude on the NV axis, $f_s$ is the oscillation frequency, and $\phi_0$ is the phase offset between the applied AC field and the dynamical decoupling sequence.
After $N_\mathrm{\pi}$ $\mathrm{\pi}$ pulses, each separated by a free precession time $\tau = 1/2f_s$, the dynamical phase accumulation by each NV in the ensemble is given by:
\begin{equation}
\label{DynamicalPhase}
\begin{aligned}
\theta(B_z,\phi_0) =&\, \frac{g\mu_B N_\mathrm{\pi}B_z}{\pi \hbar f_s}\cos\phi_0,
\end{aligned}
\end{equation}
where $g\approx2$ is the Land\'e g-factor, $\mu_B$ is the Bohr magneton, and $\hbar$ is the reduced Planck constant. 
With a $\mathrm{\pi/2}$ phase difference between the first and last $\mathrm{\pi/2}$ pulses in the sequence, the NV fluorescence readout signal is given by:
\begin{equation}
\label{DDRead}
\begin{aligned}
S(B_z,\phi_0) =& \, S_o + S_a\sin(\theta)\\
=&\, S_o + S_a\sin(\kappa B_z\cos\phi_0),
\end{aligned}
\end{equation}
where $S_o$ is the NV fluorescence offset, $S_a$ is the fluorescence oscillation amplitude, and $\kappa = g\mu_B N_\mathrm{\pi}/\pi \hbar f_s$ is a constant. 
The NV fluorescence readout corresponding to projection onto the $\ket{m_s = 0}$ and $\ket{m_s = 1}$ states are $S_o \pm S_a$, respectively. 

We calibrate the RF-QDM's performance of AC magnetometry, as follows.  
We first adjust the phase $\phi_0$ of a modest amplitude signal ($=1$\,MHz) applied to the RF test coil, by setting the AWG output amplitude to 0.01 (equivalent to 11.25$\,$mV $\mathrm{V_{pp}}$ across the RF test coil). 
We then perform AC magnetometry as described above, using the photodiode to collect the NV fluorescence with a laser beam illumination diameter $\approx50\,\mathrm{\mu m}$.
The resulting NV photoluminescence contrast can be approximated as: $S(\phi_0)\approx S_o + S_a\kappa B_z\cos\phi_0$.
By sweeping the phase offset $\phi_0$, we determine the experimental phase shift for the AC test signal, as measured with the dynamical decoupling sequence, from the value of $\phi_0$ that gives maximum NV fluorescence contrast.
As shown in [Fig. \ref{fig:7}(a)], this experimental phase shift is very small ($\sim{1^\circ}$).
Next, with the phase fixed at $\phi_0 = 0$, we sweep the AWG output amplitude (changing $B_z$ over a range comparable to that observed in microcoil experiments) to determine the AC magnetometry curve [Fig. \ref{fig:7}(b)]. 
After fitting with a sine function, we extract the AWG amplitude change $A_{2\pi}$ for $\mathrm{2\pi}$ phase accumulation by the NVs (i.e., $\theta = 2\pi$).
Based on equation (\ref{DynamicalPhase}), the AC magnetic field amplitude for $\mathrm{2\pi}$ phase accumulation $B_{2\pi}$ is given by $2\pi^2 f_s\hbar/g\mu_B N_\pi$. 
The calibration factor to convert from AWG amplitude to AC magnetic field amplitude is then calculated as $C = B_{2\pi}/A_{2\pi}$.
The derivative of the magnetometry curve at the zero-crossing point provides us with the slope of the NV fluorescence signal varying with the AWG output amplitude, $\Delta S/\Delta A$. 
For AC magnetic sensing, the zero-crossing slope is given by $\Delta S/\Delta B_z = C \Delta S/\Delta A$.
In the imaging modality, we perform this calibration process for each pixel. 
The spatial distribution of the AC magnetometry curve zero-crossing slope, which is mainly determined by the laser beam illumination of the NV layer, is shown in [Fig. \ref{fig:7}(c)].
Once an NV fluorescence image is acquired, the in-phase AC magnetic field image can be extracted using the pre-calibrated zero-crossing slope for each pixel.

\section{RF Magnetic Imaging Spatial Resolution and Frequency Range}
\label{supp:thinNV}
\begin{figure}
    \centering
    \includegraphics[width=\linewidth]{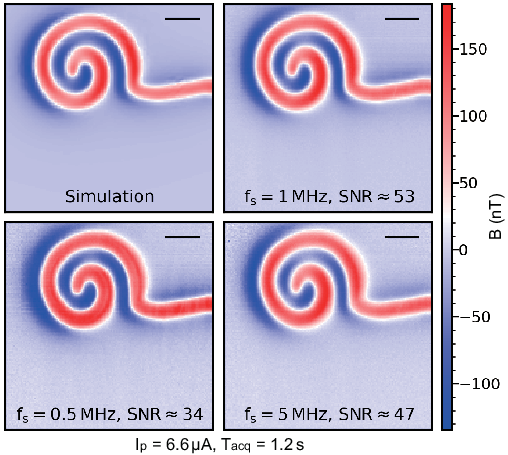}
    \caption{RF-QDM imaging of narrowband magnetic fields of known phase, using a dynamical decoupling (XY8-8) protocol and a $1.7\,\mathrm{\mu m}$-thick NV layer diamond. 1\,MHz, 0.5\,MHz, and 5\,MHz alternating currents with $I_p = 6.6\,\mathrm{\mu A}$ are applied to the upper microcoil, in separate experiments. After averaging over 1.2\,s for each signal frequency, the experimental images exhibit SNRs of approximately 50, 34, and 47, respectively, calculated using the same method as for [Fig. \ref{fig:2}(b)] in the main text. The imaged spatial patterns of RF magnetic fields are consistent over an order of magnitude of signal frequency; and also with a numerical simulation of an applied DC signal at $6.6\,\mathrm{\mu A}$. All scale bars correspond to 50$\,\mathrm{\mu m}$ in this figure. The colorbar is not centered around zero because of asymmetry in the projection of the RF magnetic field amplitude on the sensing NV axis.
    }
    \label{fig:8}
\end{figure}
To further illustrate the $\approx2\,\mathrm{\mu m}$ spatial resolution and easily tunable frequency range of the RF-QDM, we perform experiments under the same conditions as those described in [Fig. \ref{fig:2}(b)], except we use a diamond with a $1.7\,\mathrm{\mu m}$ surface NV layer. 
We begin by applying an AC signal at $1\,\mathrm{MHz}$, with a peak current of $6.6\,\mathrm{\mu A}$, to the upper microcoil on the NV-diamond surface.
After a 1.2-second acquisition (200 camera external frames), we obtain an RF magnetic field amplitude image with a peak-to-peak difference of about 300\,nT, revealing the near-field pattern generated by the upper microcoil with $\mathrm{SNR}\approx{53}$, in good agreement with a numerical simulation [Fig. \ref{fig:8}]. 
The simulation is performed using the same approach described in Sec. \ref{RFImaging} of the main text, averaging the three-dimensional magnetic field from the microcoil at a $7\,\mathrm{\mu m}$ stand-off distance over the NV layer thickness.

As mentioned in Sec. \ref{RFImaging}, the central detection frequency of the dynamical decoupling sequence is easily tunable over the MHz scale. 
The maximum sequence length, limited by the spin decoherence time $T_2$ of the NV ensemble, determines the lower limit of the signal detection frequency, $f_s$.
For the XY8-8 sequence, $T_2\approx60\,\mathrm{\mu s}$ [Fig. \ref{fig:10}], resulting in $\tau\approx 1\,\mathrm{\mu s}$ and hence a minimum $f_s\approx0.5$\,MHz. 
As a demonstration, we image the RF magnetic field generated by a 0.5\,MHz AC signal applied to the upper microcoil, with the same peak current of $6.6\,\mathrm{\mu A}$ [Fig. \ref{fig:8}]. 
Although the magnetic field pattern is consistent with the simulation and experiment at $f_s=1\,$MHz, the SNR decreases to $\approx34$, due to NV fluorescence signal decay with the longer sensing time.
To perform imaging of even lower frequency narrowband signals, the Hahn spin echo sequence can be used \cite{Pham_2011}.
For the diamonds used in the present experiment, this would allow a maximum $\tau=T_\mathrm{2,Hahn}\approx10\,\mathrm{\mu s}$, indicating that the RF-QDM is capable of imaging narrowband signals down to about 50\,kHz, albeit with reduced sensitivity.

The maximum detection frequency of the dynamical decoupling sequence is limited by the length of $\pi$ pulses, which is determined by the NV Rabi frequency of the MW driving. 
At present, the RF-QDM's on-chip $\Omega$-shaped coplanar waveguide achieves a 6.4\,MHz Rabi frequency with the maximum output of the MW amplifier ($\approx50$\,dBm). 
Since $\tau$ must be longer than the $\pi$ pulse duration, the maximum detection frequency of the RF-QDM is presently less than 6.4\,MHz.
As a demonstration, we image a 5\,MHz RF magnetic field pattern, again for $I_p = 6.6\,\mathrm{\mu A}$, with SNR $\approx47$, similar to the results obtained at 1\,MHz. 
A straightforward way to increase the Rabi frequency (and hence the maximum signal frequency that can be imaged) is by reducing the diameter of the MW antenna; however, this would worsen MW and hence Rabi frequency homogeneity, which is detrimental for wide-field imaging. 
Another option is to use resonators that provide a stronger Rabi frequency; but this approach fixes the detection frequency, limiting the versatility of the RF-QDM.
To allow imaging of higher signal frequencies ($\sim{10}\,$MHz to GHz range) without the limitations noted above, we plan in future work to investigate the use of a quantum frequency mixing protocol \cite{WangG2022} with the RF-QDM.
\section{Fourier Analysis of CASR Signal}
\label{supp:CASR}
For the CASR protocol, the phase offset of each readout is varying due to the signal being time-dependent.
The equation (\ref{DDRead}) can be rewritten as:
\begin{equation}
\label{DDReadT}
\begin{aligned}
S(t,B_z,\phi_0) =& \,S_o+ S_a\sin[\kappa B_z \cos(2\pi f_s t+\phi_0)]\\
                =&\, S_o+2S_a\sum^\infty_{n=0}(-1)^nJ_{2n+1}(\kappa B_z)\\
                 &\,\cdot\cos[(2n+1)(2\pi f_s t+\phi_0)],
\end{aligned}
\end{equation}
where $J_n(x)$ is the n-th order Bessel function and $f_s$ is the signal central frequency. The CASR spectrum after an FFT is given by:
\begin{equation}
\label{SFFT}
\begin{aligned}
&\,S(f,B_z,\phi_0) = \sqrt{2\pi}\{S_o\delta(f) + 2\pi S_a\sum^\infty_{n=0}(-1)^nJ_{2n+1}(\kappa B_z)\\
&\,\cdot[e^{-i\phi_0}\delta(f-(2n+1)f_s)+e^{i\phi_0}\delta(f+(2n+1)f_s)]\}\\
&\,= S_\mathrm{DC}+S_\mathrm{Real}+iS_\mathrm{Imag},
\end{aligned}
\end{equation}
where $\delta(f)$ is the Dirac function in frequency space, and $S_\mathrm{DC}$, $S_\mathrm{Real}$, and $S_\mathrm{Imag}$ are the DC component, real part, and imaginary part of the Fourier spectrum, respectively. $S_\mathrm{DC} =\sqrt{2\pi}S_o\delta(f)$, whereas the real and imaginary parts are given by:
\begin{equation}
\label{FFTRI}
\begin{aligned}
&\,S_\mathrm{Real}(f,B_z,\phi_0)= (2\pi)^{\frac{3}{2}}S_a\cos(\phi_0)\sum^\infty_{n=0}(-1)^nJ_{2n+1}(\kappa B_z)\\
&\,\cdot[\delta(f+(2n+1)f_s)+\delta(f-(2n+1)f_s)],\\
&\,S_\mathrm{Imag}(f,B_z,\phi_0)= (2\pi)^{\frac{3}{2}}S_a\sin(\phi_0)\sum^\infty_{n=0}(-1)^nJ_{2n+1}(\kappa B_z)\\
&\,\cdot[\delta(f+(2n+1)f_s)-\delta(f-(2n+1)f_s)].
\end{aligned}
\end{equation}
In RF-QDM CASR experiments reported in Sec. \ref{CASRImaging}, we utilize the XY8-8 sequence for imaging RF magnetic field patterns with $B_{z,max}\approx40\,\mathrm{nT}$, resulting in $\kappa B_{z,max}\approx0.8$.
In the regime of $\kappa B_z < 1$, we can ignore the contribution of higher order Bessel functions. 
Keeping the first order Bessel term's positive part due to the symmetry of the Fourier spectrum, the real and imaginary parts can be simplified as:
\begin{equation}
\label{FFTRIP}
\begin{aligned}
S^+_\mathrm{Real}(f,B_z,\phi_0)=&\, (2\pi)^{\frac{3}{2}}S_a\cos(\phi_0)J_1(\kappa B_z)\delta(f-f_s),\\
S^+_\mathrm{Imag}(f,B_z,\phi_0)=&\, -(2\pi)^{\frac{3}{2}}S_a\sin(\phi_0)J_1(\kappa B_z)\delta(f-f_s).
\end{aligned}
\end{equation}
For RF-QDM CASR experiments, the measurement separation is $t_{m} = N_{d}\cdot t_\mathrm{Seq}$, where $t_\mathrm{Seq}$ is the sequence length for one camera demodulation cycle. The camera frame rate $f_\mathrm{SR} = 1/t_{m}$. 
Based on the Nyquist–Shannon sampling theorem, the bandwidth of the measured Fourier spectrum is $f_\mathrm{SR}/2$. 
The CASR alias frequency is $f_a =\, \mid f_s - mf_\mathrm{SR}\mid$, where $m$ is the integer closest to $f_s/f_\mathrm{SR}$. 
Consequently, for each pixel in a CASR image, the measured signal components are:
\begin{equation}
\label{FFTRIPA}
\begin{aligned}
S^{+}_\mathrm{Real,m}(f,B_z,\phi_0)=&\, (2\pi)^{\frac{3}{2}}S_a\cos(\phi_0)J_1(\kappa B_z)\delta(f-f_a),\\
S^{+}_\mathrm{Imag,m}(f,B_z,\phi_0)=&\, -(2\pi)^{\frac{3}{2}}S_a\sin(\phi_0)J_1(\kappa B_z)\delta(f-f_a).
\end{aligned}
\end{equation}
Here, we define the amplitude to have the same sign as the real component; thus the FFT spectrum amplitude of a measured RF magnetic field is given by:
\begin{equation}
\label{FFTAmp}
\begin{aligned}
S^{+}_\mathrm{Amp,m}(f,B_z)=&\, \frac{S^{+}_\mathrm{Real,m}}{\mid S^{+}_\mathrm{Real,m}\mid}\cdot \sqrt{{S^{+}_\mathrm{Real,m}}^2 + {S^{+}_\mathrm{Imag,m}}^2}\\
=&\,(2\pi)^{\frac{3}{2}}S_a J_1(\kappa B_z)\delta(f-f_a).
\end{aligned}
\end{equation}
By integrating the relevant peak of the FFT spectrum, the CASR signal at each frequency component is:
\begin{equation}
\label{SCASR}
\begin{aligned}
S_\mathrm{CASR}(B_z) = &\, \int_{f_a-\Delta f}^{f_a+\Delta f}S^{+}_\mathrm{Amp,m}(f,B_z)\,df\\
=&\,(2\pi)^{\frac{3}{2}}S_a J_1(\kappa B_z).
\end{aligned}
\end{equation}
In our experiment, the peak half linewidth $\Delta f\approx1\,$Hz, resulting in $S_\mathrm{CASR}\approx S^{+}_\mathrm{Amp,m}$. 
When $\kappa B_z < 1$, the first order Bessel function closely approximates a linear function, giving $S_\mathrm{CASR}\propto B_z$.
In the imaging modality, the NV fluorescence signal amplitude $S_a$ has a spatial distribution due to the laser beam shape. 
Thus, we normalize the FFT image with the NV fluorescence image [Fig. \ref{fig:6}(b)] to obtain the final CASR image of the RF magnetic field amplitude.

For a CASR phase image, we calculate the RF magnetic field phase offset for each pixel using the real and imaginary components of the Fourier spectrum, given by equation (\ref{FFTRIPA}), as follows:
\begin{equation}
\label{CASRPhase}
\begin{aligned}
\phi_0 =&\,\arctan{(-S^{+}_\mathrm{Imag,m}/S^{+}_\mathrm{Real,m})}.
\end{aligned}
\end{equation}
We then construct CASR phase images at each alias frequency within the spectral bandwidth.

\section{CASR Imaging Experimental Details}
\label{supp:CASRDetails}
\begin{figure}
    \centering
    \includegraphics[width=\linewidth]{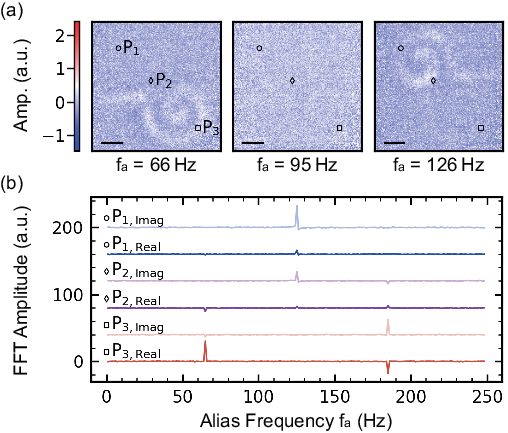}
    \caption{(a) CASR images of RF magnetic field amplitudes, for the same experimental parameters as in [Fig. \ref{fig:4}] of the main text, with alias frequency values of 66\,Hz and 126\,Hz, offset by 1\,Hz from the applied signal frequency components (tones) at 65\,Hz and 125\,Hz, respectively; and at an intermediate alias frequency of 95 Hz. All scale bars correspond to 50$\,\mathrm{\mu m}$. The colorbar is not centered around zero because of asymmetry in the projection of the RF magnetic field amplitude on the sensing NV axis. (b) Real and imaginary components of the CASR spectrum, as a function of alias frequency, at three test pixels $\mathrm{P_{1,2,3}}$ with positions labeled in (a). Peaks corresponding to the three applied signal frequency components (tones) are illustrated in the phase offset polar plot in [Fig. \ref{fig:4}(d)].}
    \label{fig:9}
\end{figure}

As noted in Sec. \ref{CASRImaging} of the main text, CASR imaging data is acquired with a 500\,Hz frame rate for a 1-second measurement. 
To improve image SNR, we perform a 1-second background measurement (with no RF signal applied to the microcoils) after each CASR image acquisition (with signal applied); and repeat 100 times for signal averaging. 
During the imaging process, the camera saves 500 external frames in its RAM, transfers it to the computer, and clears the buffer to start the next iteration of the experiment. 
The total measurement time is 200 seconds; however, because of the data transfer time, the wall time for the experiment is approximately 30 minutes. 
The total data size is large (17\,GB); but after background subtraction and averaging the CASR signal is in the time domain for each pixel, resulting in a $150\times150\times500$ matrix, with a data size of only 80\,MB. 
The computational time of the Fourier transformation for all pixels across the FOV is 0.4 seconds.

The current implemented generation of Heliotes cameras is unable to read and write data simultaneously. 
Instead, it temporarily stores the measurement data in its memory buffer. Once the buffer is nearly full, it pauses the experiment to transfer the data to the computer and clear the buffer, which takes approximately 5 seconds. 
This limitation could be addressed by using cameras capable of directly saving data to the computer, which would significantly improve performance for practical applications of the RF-QDM.

In the main text, we show CASR images at each applied RF signal frequency component (tone), and also at alias frequencies far from any signal tone [Fig. \ref{fig:4}(a)]. 
In [Fig. \ref{fig:9}(a)], we provide complementary CASR amplitude images, for the same experimental parameters, at alias frequencies with a small offset (1\,Hz) from the signal tones at 65\,Hz and 125\,Hz; and at an intermediate alias frequency (95\,Hz).
The colorbar range is set 20 times smaller than in [Fig. 4(a)] to visualize the weak signal amplitudes when detuned even by 1\,Hz from the applied signal tones.
In [Fig. \ref{fig:9}(b)], we show $S^{+}_\mathrm{Real,m}$ and $-S^{+}_\mathrm{Imag,m}$ as a function of alias frequency for the three test pixels $\mathrm{P_{1,2,3}}$. 
These results complement the polar plot of [Fig. \ref{fig:4}(d)], which depicts the phase offsets and peak values of the real and imaginary components of the CASR spectrum, for each signal frequency component. 
The phase error in our CASR measurements is estimated as:

\begin{equation}
\label{PhaseError}
\begin{aligned}
\delta\phi =\arcsin{\left(\frac{\sigma_1}{S_\mathrm{CASR}(f_2)}\right)}\approx0.6^\circ,
\end{aligned}
\end{equation}

where $\sigma_1$ is the noise floor of the CASR spectrum of the test pixel $\mathrm{P_1}$ in [Fig. \ref{fig:4}(b)], and $S_\mathrm{CASR}(f_2)$ is the CASR amplitude of the alias frequency at $f_2 = 125$ Hz. 

\section{AC Magnetic Sensitivity Optimization}
\label{supp:ACSensOpt}
\begin{figure}
    \centering
    \includegraphics[width=\linewidth]{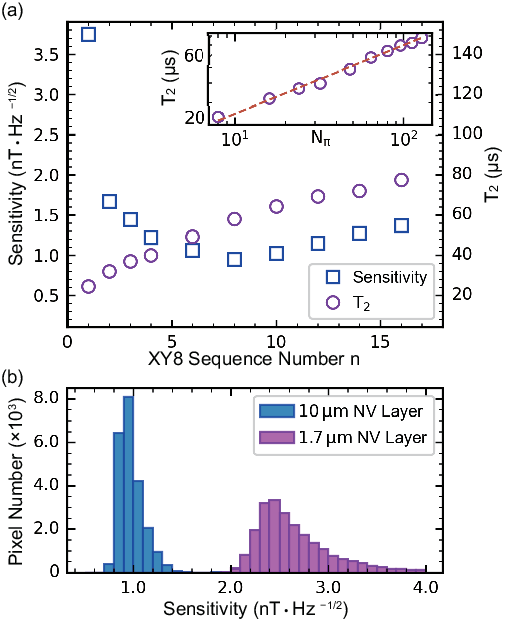}
    \caption{(a)AC magnetic sensitivity optimization with varying XY8 sequence number, performed with photodiode measurements of the 10$\,\mathrm{\mu m}$ NV layer diamond. 
    As the XY8-n sequence number increases, the NV spin coherence time $T_2$ increases.
    The optimized median single-pixel sensitivity across the FOV is achieved with XY8-8. 
    The inset illustrates the fitting of the measured coherence time $T_2$ with $\pi$ pulse number $N_\mathrm{\pi} = 8n$, showing a power law. 
    The fitted slope $s$ = 0.44(1).
    (b) RF-QDM imaging using the lock-in camera and a 1\,MHz signal applied with the RF test coil. 
    Histogram showing the distribution of single-pixel AC sensitivity across the FOV using the XY8-8 sequence. 
    The median sensitivity is $0.93(1)\,$nT$\cdot$Hz$^{-1/2}$ for the 10$\,\mathrm{\mu m}$-thick NV layer diamond, and $2.56(1)\,$nT$\cdot$Hz$^{-1/2}$ for the 1.7$\,\mathrm{\mu m}$-thick NV layer diamond.
    } 
    \label{fig:10}
\end{figure}
For an AC magnetic field measurement by an NV sensor using a dynamical decoupling sequence, the magnetic sensitivity limited by photon shot noise and spin projection noise is approximately given by \cite{Barry2020}:
\begin{equation}
\label{eq:sensitivity}
\begin{aligned}
\eta_\mathrm{AC} = \frac{\pi}{2\Delta m_s \gamma} \frac{1}{Ce^{(-t_\mathrm{S}/T_2)^p}\sqrt{N}}\frac{\sqrt{t_\mathrm{I,R}+t_\mathrm{S}}}{t_\mathrm{S}},
\end{aligned}
\end{equation}
where $\Delta m_s = 1$ accounts for the difference of $m_s$ states used for sensing, $C$ is the optical measurement contrast at sensing time $t_\mathrm{S}=0$, $T_2$ is the NV spin coherence time, $p$ describes the NV spin coherence envelope decay shape \cite{Bauch2020}, $N$ is the average photon number collected per measurement, and $t_{I,R}$ is the combined initialization and readout time during the dynamical decoupling sequence.
The spin coherence time measured with dynamical decoupling sequences scales with a power law as a function of the number of $\pi$ pulses \cite{Pham2012}, $T_2(N_\pi) = {T_2(1)N_\pi}^s$, where s is determined by the noise spectrum of the decohering spin bath and typically exhibits sublinear behavior \cite{deSousa2009}.
For a given AC signal at frequency $f_s$, the sensing time has a linear dependence on the $\pi$ pulse number $t_\mathrm{S} = 2N_\pi/f_s$.
After experimental adjustment of $C$, $N$ and $t_{I,R}$, the AC sensitivity is optimized with a specific number of pulses \cite{Barry2020}:
\begin{equation}
\label{eq:pipulsenum}
\begin{aligned}
N_{\pi,opt}=\biggl[ \frac{(2f_sT_2(1))^p}{2p(1-s)}\biggr]^{\frac{1}{p(1-s)}}.
\end{aligned}
\end{equation}
In photodiode measurements with the 10 µm NV layer diamond, we determine a typical, spatially-averaged value of $T_2(1) = 10\,\mathrm{\mu s}$ using a Hahn spin echo sequence \cite{PhysRev.80.580} and $p=1.3$. 
By fitting the measured coherence time variation with pulse number, we derive a value of $s = 0.44(1)$ [Fig. \ref{fig:10}(a) inset]. 
Using these parameters, for $f_s = 1\,$MHz we estimate $N_{\pi,opt}/8$ to be approximately 16, indicating that the XY8-16 sequence provides optimized sensitivity. 
However, during RF-QDM magnetic field imaging with the lock-in camera, the contribution of camera quantization noise to AC magnetic sensitivity must also be considered.
With a fixed number of demodulation cycles, camera noise increases with sensing sequence length, as shown in Appendix \ref{supp:Noise} below.
Considering both shot noise and camera noise, the optimized sensing sequence should be shorter than XY8-16.

To ascertain the optimal sensing sequence including camera noise, we vary the number of XY8 sequences and measure the AC sensitivity of each pixel across the FOV. 
The experimental per-pixel magnetic sensitivity results are computed from $\eta=\delta B\sqrt{T_{acq}}$, where $\delta B$ is the smallest magnetic field that can be measured with $\mathrm{SNR}=1$ after acquisition time $T_{acq}$.
For a camera operating at an external frame rate $f_\mathrm{SR}$, a series of magnetic noise images are collected. 
Then the magnetic noise of each pixel, $\sigma_{pxl}$, is obtained from the standard deviation of the magnetometry data from that pixel across all recorded frames. 
The measured per-pixel magnetic sensitivity is then given by:
\begin{equation}
\label{eq:sensitivitypractical}
\begin{aligned}
\eta =  \frac{\sigma_{pxl}}{\sqrt{f_\mathrm{SR}}}.
\end{aligned}
\end{equation}
The results displayed in [Fig. \ref{fig:10}(a)] are obtained using 500 frames collected for $f_\mathrm{SR}=1/[4N_d(t_\mathrm{I,R}+t_\mathrm{S})]$, where demodulation cycle number $N_d = 34$,  $t_\mathrm{I,R} = 26\,\mathrm{\mu s}$, $t_\mathrm{S} = N_\pi/(2f_s)$, and the signal frequency $f_s = 1\,$MHz (applied with the RF test coil).
The optimized sequence is determined  to be XY8-8, with a median per-pixel AC sensitivity of $0.93(1)\,$nT$\cdot$Hz$^{-1/2}$.
Given the sensing voxel volume of $40\,\mathrm{\mu m}^3$, we calculate a volume-normalized AC sensitivity of $5.88$\,nT$\cdot$Hz$^{-1/2}\cdot\mathrm{\mu m}^{3/2}$, which is comparable to state-of-the-art photodiode-based XY8 sensitivity for AC magnetometry using NV ensembles \cite{Arunkumar2023}.
We also perform an analogous characterization and AC magnetometry optimization for the 1.7$\,\mathrm{\mu m}$-thick NV layer diamond with similar results.
In particular, we determine a median per-pixel AC sensitivity of $2.56(1)\,$nT$\cdot$Hz$^{-1/2}$ at the same experimental condition [Fig. \ref{fig:10}(b)].
The volume-normalized AC sensitivity is calculated to be $6.68$\,nT$\cdot$Hz$^{-1/2}\cdot\mathrm{\mu m}^{3/2}$, which is slightly worse than for the 10$\,\mathrm{\mu m}$ NV layer diamond.
The reason for this difference is that the camera quantization noise (discussed in Appendix \ref{supp:Noise}) is larger for the 1.7$\,\mathrm{\mu m}$ NV layer diamond, due to this diamond's fewer number of NVs in the thinner layer, resulting in a lower total NV fluorescence rate per imaging pixel, and hence a reduced magnetometry slope.

\section{Noise Source Analysis}
\label{supp:Noise}
For RF-QDM measurements presented here, photon shot noise and camera quantization noise are the major limitations to AC magnetic sensitivity.
We estimate shot-noise-limited AC sensitivity using equation (\ref{eq:sensitivity}), introduced above.
By focusing the NV fluorescence with a laser beam illumination diameter $\approx50\,\mathrm{\mu m}$ onto a photodiode, we experimentally determine typical NV-ensemble-averaged XY8-8 parameters for the 10$\,\mathrm{\mu m}$ NV layer diamond to be about $C = 3$\,\%, $T_2 = 58\,\mathrm{\mu s}$, and $p=1.3$. 
As mentioned in the previous section, we use $t_\mathrm{I,R} = 26\,\mathrm{\mu s}$ and $t_\mathrm{S} = 8\times8\times0.5 = 32\,\mathrm{\mu s}$ for a XY8-8 RF-QDM magnetic imaging experiment, and estimate $N=4\times 10^4$ photons are collected in a single camera pixel \cite{Tang2023}.
With these values, the estimated photon shot-noise-limited per-pixel AC magnetic sensitivity for a 1\,MHz signal frequency is: $\eta_\mathrm{shot} = 0.3\,$nT$\cdot$Hz$^{-1/2}$.

The quantization noise of the Heliotis lock-in camera includes electronic and digitization noise. For each external frame, the standard deviation of the noise is dependent on the number of demodulation cycles ($N_d$) spent accumulating differences between successive internal exposures, and can be estimated in the camera device unit (DU) as \cite{heliotis}
\begin{equation}
\label{suppeq:quantizationnoise}
\sigma_\mathrm{quant} = \sqrt{0.81+0.16N_d}.
\end{equation}  
To convert the DU estimate of quantization noise to equivalent magnetic noise, we calculate the median zero-crossing slope of imaged AC magnetometry curves across the whole FOV [Fig. \ref{fig:7}(c)] and find $\overline{\Delta S/\Delta B_z} = 258$\,DU/$\mathrm{\mu T}$. 
The estimated quantization-noise-limited per-pixel AC magnetic sensitivity is given by:
\begin{equation}
\label{suppeq:quantizationsensitivity}
\eta_\mathrm{quant} = \frac{\sigma_\mathrm{quant}}{\overline{\Delta S/\Delta B_z}\sqrt{f_\mathrm{SR}}},
\end{equation}
which is about $0.86\,$nT$\cdot$Hz$^{-1/2}$ for the RF-QDM using the 10$\,\mathrm{\mu m}$ NV layer diamond.
In the process of optimizing camera quantization noise, increasing the demodulation cycle number $N_d$ leads to a larger $\sigma_\mathrm{quant}$ and a smaller sample rate $f_\mathrm{SR}$. 
However, when operating in the regime where the photon count for each pixel does not saturate the well depth of the camera, the magnetometry slope $\overline{\Delta S/\Delta B_z}$ increases linearly with $N_d$, leading to less quantization noise. 
Experimentally, we sweep the demodulation cycle number $N_d$ and observe a linear region for $N_d\leqslant34$. 
Then, we use the maximum $N_d$ to utilize the full digitizer range of the camera in the experiment.
After adding photon shot noise and camera quantization noise in quadrature, the combined estimate of per-pixel AC magnetic sensitivity is $\eta_\mathrm{est} = \sqrt{\eta_\mathrm{shot}^2 + \eta_\mathrm{quant}^2} = 0.91\,$nT$\cdot$Hz$^{-1/2}$, which is consistent with the measured median per-pixel magnetic sensitivity of $0.93(1)\,$nT$\cdot$Hz$^{-1/2}$. 

For the present RF-QDM, the AC magnetic field sensitivity is limited by the Heliotis lock-in camera. 
The dynamic range of the Heliotis camera is fixed (DU from 0 to 1023, 10-bit), and the camera sensor is optimized for unity gain \cite{heliotis}. 
For a given number of demodulation cycles $N_d$, the quantization noise remains constant. 
Thus, enhancing the median zero-crossing slope of imaged AC magnetometry curves, which is utilized for converting the camera external frame output from DU to a magnetic-field value, can effectively decrease the magnetic-noise contribution from quantization noise.
The slope of an AC magnetometry curve for a given $N_d$ is directly proportional to both NV fluorescence intensity and spin-state readout contrast, factors that influence the accumulated differential photon number $\Delta N$ in a camera external frame. 
Moreover, the slope also depends on the NV spin coherence time $T_2$. 
Consequently, RF-QDM sensitivity will benefit from efforts to increase NV fluorescence intensity, spin-state readout contrast, and $T_2$; as well as diamond engineering efforts to enable increased laser illumination intensity while mitigating contrast degradation due to NV charge-state conversion \cite{edmonds2021characterisation}.

We note an alternative approach for wide-field NV-diamond measurements using an array of single-photon avalanche diodes (SPADs) \cite{Wang2023}.  The SPAD array provides single-photon counting capability and high time resolution (frame rate up to 100\,kHz).  However, it is only applicable to low photon counts, with correspondingly greatly reduced magnetic field sensitivity per pixel compared to a camera-based QDM.  The SPAD array is thus best suited for very-thin ($<100$\,nm) NV layers with single/few NVs per pixel.
\bibliography{references}

\end{document}